\documentclass[aps,prl,twocolumn,showpacs,amsmath,amssymb,superscriptaddress,longbibliography,floatfix]{revtex4-2}

\usepackage[pdftex]{graphicx}
\usepackage[breaklinks,colorlinks=true,citecolor=blue]{hyperref}
\usepackage{braket}
\usepackage{lipsum}
\usepackage{xcolor}
\usepackage{siunitx}

\usepackage[capitalize]{cleveref} 

\newcommand{\md}{\mathrm{d}}
\newcommand{\mi}{\mathrm{i}}
\newcommand{\me}{\mathrm{e}}
\definecolor{crimson}{RGB}{255,102,255}

\usepackage{sidecap,tikz}
\definecolor{lime}{HTML}{A6CE39}
\DeclareRobustCommand{\orcidicon}{\hspace{-1mm}
	\begin{tikzpicture}
		\draw[lime, fill=lime] (0,0) 
		circle [radius=0.16] 
		node[white] {{\fontfamily{qag}\selectfont \tiny \,ID}};
		\draw[white, fill=white] (-0.0525,0.095) 
		circle [radius=0.007];
	\end{tikzpicture}
	\hspace{-3mm}
}
\foreach \x in {A, ..., Z}{\expandafter\xdef\csname orcid\x\endcsname{\noexpand\href{https://orcid.org/\csname orcidauthor\x\endcsname}
		{\noexpand\orcidicon}}
}

\begin{document}
	
\title{On-Demand and Tunable Andreev-Conversion of Single-Electron Charge Pulses}

\author{Pablo Burset\orcidA{}}
\affiliation{Department of Theoretical Condensed Matter Physics\char`,~Universidad Aut\'onoma de Madrid, 28049 Madrid, Spain}
\affiliation{Condensed Matter Physics Center (IFIMAC), Universidad Aut\'onoma de Madrid, 28049 Madrid, Spain}
\affiliation{Instituto Nicol\'as Cabrera, Universidad Aut\'onoma de Madrid, 28049 Madrid, Spain}

\author{Benjamin Roussel\orcidB{}}
\affiliation{Department of Applied Physics,
	Aalto University, 00076 Aalto, Finland}

\author{Michael Moskalets\orcidC{}}
\affiliation{Department of Metal and Semiconductor Physics\char`,~NTU “Kharkiv Polytechnic Institute”, 61002 Kharkiv, Ukraine}
\affiliation{Institute for Cross-Disciplinary Physics and Complex Systems 
	IFISC (UIB-CSIC), 07122 Palma de Mallorca, Spain}

\author{Christian Flindt\orcidD{}}
\affiliation{Department of Applied Physics,
	Aalto University, 00076 Aalto, Finland}

\date{\today}

\begin{abstract} 
Electron quantum optics explores coherent single-electron charge pulse propagation in electronic nanoscale circuits akin to table-top photon setups. 
While past experiments focused on normal-state conductors, incorporating superconductors holds promise for exploiting the electron-hole degree of freedom in quantum sensing applications and quantum information processing. 
Here, we propose and analyze an on-demand and tunable mechanism for converting single-electron pulses into holes through Andreev processes on a superconductor. We develop a Floquet-Nambu scattering formalism to demonstrate the dynamic conversion of charge pulses and the controllable generation of coherent electron-hole superpositions through interferometric magnetic flux control based on the chiral edge states of a quantum Hall sample. Our discussion covers optimal conditions in realistic scenarios, affirming the feasibility of our proposal with current technology.
\end{abstract}

\maketitle

{\it Introduction.---} Recent advances in dynamic quantum transport have paved the way for coherent single-electron control and manipulation in nanoscale circuits~\cite{Erwann_AdP,Janine_PSS,Waintal_RPP,Weinbub_2022}. Single electrons can now be emitted into a coherent conductor without disturbing the underlying Fermi sea by applying Lorentzian voltage pulses to an ohmic contact~\cite{Levitov_1996,Levitov_1997,Safi1999,Levitov_2006,Dubois_2013,Jullien_2014,Assouline:2023,Chakraborti2025}. Moreover, by coupling single-electron emitters to the chiral edge states of a quantum Hall sample, electronic interferometers can be realized that are similar to those from quantum optics~\cite{Feve_2007,Bocquillon_2012,Bocquillon_2013,Kataoka_2013,Ubbelohde_2014,Kataoka_2015,Glattli_2016,Roulleau_2021,Ubbelohde:2023,Fletcher:2023,Wang:2023}. This growing field of research has been dubbed electron quantum optics as it borrows ideas and concepts from the quantum theory of light. However, despite many similarities, there are also marked differences between photons traveling in a waveguide and electrons propagating on top of the Fermi sea in an electronic circuit. For example, due to their fermionic nature, electrons arriving simultaneously at a beam splitter tend to anti-bunch~\cite{Bocquillon_2013}, unlike photons that rather bunch and exit via the same output arm~\cite{Hong_1987}. 

Another important difference is the existence of holes in the Fermi sea, which play the role of antiparticles for electrons, and which have no counterparts in quantum optics. This additional degree of freedom opens up a wide range of possibilities, such as the production of charge-neutral heat pulses~\cite{Portugal:2023} and the generation of electron-hole entanglement~\cite{Dasenbrook_2015,Chen2022,Martin_2023,Chen2025}. However, to fully exploit the electron-hole degree of freedom, a controllable mechanism is needed to produce superpositions of electrons and holes on demand. 
Andreev reflections on a superconductor seem ideal for this purpose, and recently quantum Hall conductors have been connected to superconductors~\cite{Zulicke_2005,Ustinov_2007,Schonenberger_2012,Rokhinson_2015,Calado2015,BenShalom2016,Ren_2017,Das_2018,Xu_2019,Finkelstein_2020}, showing signatures of electron-hole conversion~\cite{Lee_2017,Das_2021,Kim2022,Shabani_2022,Finkelstein_2023,Uday2023} and supercurrents mediated by quantum Hall edge states over micrometers in experiments with static voltages~\cite{Vignaud2023,Amet_2016}. Despite these developments, superconductors have not yet been incorporated as a building block in experiments on electron quantum optics with dynamic single-electron emitters. 

\begin{figure}[b!]
\includegraphics[width=1.\columnwidth]{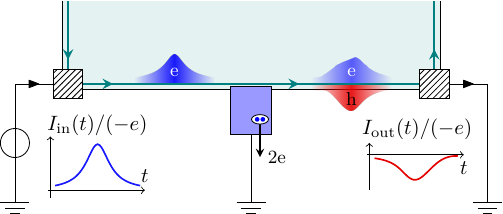}
\caption{\label{fig:setup} 
		Andreev conversion of a charge pulse. Clean single-electron states are injected into a chiral edge state by applying Lorentzian-shaped voltage pulses to the input contact. Through partial Andreev reflections on a superconductor, the charge-pulses are converted into coherent superpositions of an electron (e) and a hole (h). The currents before and after the superconductor are shown in blue and red, respectively.
	}
\end{figure} 

In this Letter, we propose and analyze a tunable setup for the coherent on-demand conversion of electrons into holes by emitting single-electron charge pulses onto a superconductor. \Cref{fig:setup} illustrates a chiral quantum Hall edge state connected to a superconductor, together with the current in the outgoing edge channel resulting from Andreev conversions of the incoming charge pulses. This setup, where current only flows in one direction due to the chiral nature of the edge states, provides the basis for the tunable electron-hole converter that we propose, which realistically can be implemented based on recent experiments~\cite{Kim2022,Shabani_2022,Finkelstein_2023}. 
Below, we present a self-contained discussion of the Floquet-Nambu scattering theory that we develop for calculating the output currents after the superconductor with the full technical details deferred to a companion paper~\cite{Burset_long}. To keep the discussion simple, we here consider spin-singlet superconductors and refer the reader to Ref.~\cite{Burset_long} for an in-depth discussion of triplet pairings and other extensions of our formalism, as well as to Refs.~\cite{Tien_1963,Vanevic_2015,Martin_2019,Sassetti_2020,Vasenko_2020,Martin_2023c} for other examples of time-dependent transport in combination with superconductors. 

{\it Floquet--Nambu scattering.---} The edge state in \cref{fig:setup} functions as a waveguide for incoming electrons. Clean single-electron excitations are injected into the circuit by applying time-dependent voltage pulses to the input. We aim to account for the scattering of the charge pulses on the superconductor connected to the edge channel. Equilibrium particles are described by the creation operators $\boldsymbol{a}^{\dagger}_{\sigma}(E)$, where $E$ is the excitation energy measured with respect to the chemical potential of the superconductor, and $\sigma=\uparrow,\downarrow$ labels the spin~\footnote{
With $\nu=2$ edge states, the spin $\sigma$ also labels the edge states. Thus, in a spin-flip process electrons are scattered from one (incoming) channel to the other (outgoing) one.
}. Within the wideband approximation, all particles propagate with the Fermi velocity $v_F$, since the dispersion relation can be linearized as $E\simeq \hbar v_F(k-k_F)$ close to the Fermi momentum $k_F$.

Superconducting correlations couple electrons with opposite energies and spins (spin-singlet states) as expressed by a Nambu spinor, $\hat{\boldsymbol{a}}_\sigma(E) = [ \boldsymbol{a}_{e\sigma}(E) , \boldsymbol{a}_{h\bar{\sigma}}(E) ]^{T}$~\footnote{
We only consider spin-singlet pairing on $\nu=2$ edge states. Spin degeneracy allows us to decouple the full spin-Nambu basis into the reduced Nambu spinors $\hat{\boldsymbol{a}}_\sigma(E)$. See Ref.~\cite{Burset_long} for more details. 
}, having used that the annihilation of a hole corresponds to the creation of an electron, i.e., $\boldsymbol{a}_{e\sigma}(E) = \boldsymbol{a}_{\uparrow,\downarrow}(E)$ and $\boldsymbol{a}_{h\bar{\sigma}}(E) = \boldsymbol{a}^\dagger_{\downarrow,\uparrow}(-E)$. 
The operators $\boldsymbol{a}_{\nu\sigma}(E)$ then correspond to electron ($\nu=e$) and hole-like ($\nu=h$) quasiparticles at equilibrium and fulfill 
$\langle \boldsymbol{a}^\dagger_{\nu\sigma}(E) \boldsymbol{a}_{\nu'\sigma'}(E')\rangle = \delta_{\nu\nu'}\delta_{\sigma\sigma'}\delta(E-E')f(E)$, where $f(E) = 1/(1+\me^{E/k_BT})$ is the Fermi function of the reservoirs at zero voltage and temperature $T$. 

The transmission of electrons can be accounted for by a Floquet scattering matrix, $\hat{S}_F(E_{n},E)$, whose elements are the probability amplitudes for a particle to scatter off the superconductor after having exchanged~$n$ energy quanta of size $\hbar\Omega$ with the driving field and changed its energy from $E$ to $E_{n}= E+n\hbar\Omega$, where $\Omega = 2\pi/\mathcal{T}$ is the frequency of the drive~\cite{Moskalets_2002}. 
The scattering amplitudes relate the electron operators in second quantization for incoming and outgoing excitations, $\hat{\boldsymbol{a}}$ and $\hat{\boldsymbol{b}}$, as
\begin{equation}\label{eq:scat-ops}
\hat{\boldsymbol{b}}(E) = \sum\limits_n \hat{S}_F(E,E_{n})\hat{\boldsymbol{a}}(E_{n}) ,
\end{equation}
having defined the Floquet-Nambu scattering matrix
\begin{equation}\label{eq:scat-mat}
\hat{S}_F(E,E_{n})= \left(\begin{array}{cc}
S_{ee}(E,E_{n}) & S_{eh}(E,E_{n}) \\
S_{he}(E,E_{n}) & S_{hh}(E,E_{n})
\end{array}\right),
\end{equation}
using hats for spinors and matrices in Nambu space. The field operator in the outgoing lead is given by
\begin{equation}\label{eq:field-op}
\hat{\boldsymbol{\Psi}}(x,t)=\frac{1}{\sqrt{h v_F}} \int_{-\infty}^{\infty} \!\mathrm{d}E \, \me^{-\mi Et/\hbar} \hat{\phi}_E(x) \hat{\boldsymbol{b}}(E) ,
\end{equation}
where $\hat{\phi}_E(x) \!=\! \mathrm{diag}_{N}(\me^{-\mi k_e(E)x},\me^{\mi k_h(E)x})$ is diagonal in Nambu space and $k_{e,h}(E)= k_F\pm E/(\hbar v_F)$. 

{\it Excess correlation function \& average current.---} The transport properties of the quasiparticles that are excited by the voltage pulses are encoded in the first-order correlation function~\cite{Degiovanni_2011,Haack_2013,Moskalets_2015,Haack_2016,Moskalets_2016,Moskalets_2020,Kotilahti_2021}. To account for the superconducting correlations, we define the Floquet-Nambu correlation function for the scattered particles as
\begin{equation}\label{eq:correlator}
\hat{\mathcal{G}} (x,t;x',t') = \langle\hat{\boldsymbol{\Psi}}^{\dagger}(x',t') \otimes \hat{\boldsymbol{\Psi}}(x,t)\rangle ,
\end{equation}
where $\otimes$ is a tensor product in Nambu space of the field operators in \cref{eq:field-op} and the quantum average is taken with respect to the equilibrium state. The correlation function is additive in the number of electrons, making it useful to characterize single-particle excitations. We thus compare the correlation function with (on) and without (off) the drive and define the excess correlation function, 
\begin{equation}\label{eq:excess}
\hat{G} (x,t;x',t') = \hat{\mathcal{G}}_\text{on}(x,t;x',t') - \hat{\mathcal{G}}_\text{off}(x,t;x',t') ,
\end{equation}
which also yields the time-dependent current as
\begin{equation}\label{eq:current}
I(t) = -g_s e v_F\mathrm{Tr}_{N} \left\{\hat{G} (x,t;x,t) \hat{\tau}_z \right\}/2 ,
\end{equation}
where $\hat{\tau}_{x,y,z}$ are the Pauli matrices acting on the Nambu space, $\mathrm{Tr}_{N}$ denotes the trace in Nambu space, and we have included the spin degeneracy $g_s=2$ for singlet pairing. Due to the linear dispersion, we can set $x'=x$ at an arbitrary position in the output lead, and we then have all the necessary ingredients to calculate the current. 

\begin{figure*}
	\includegraphics[width=1.\textwidth]{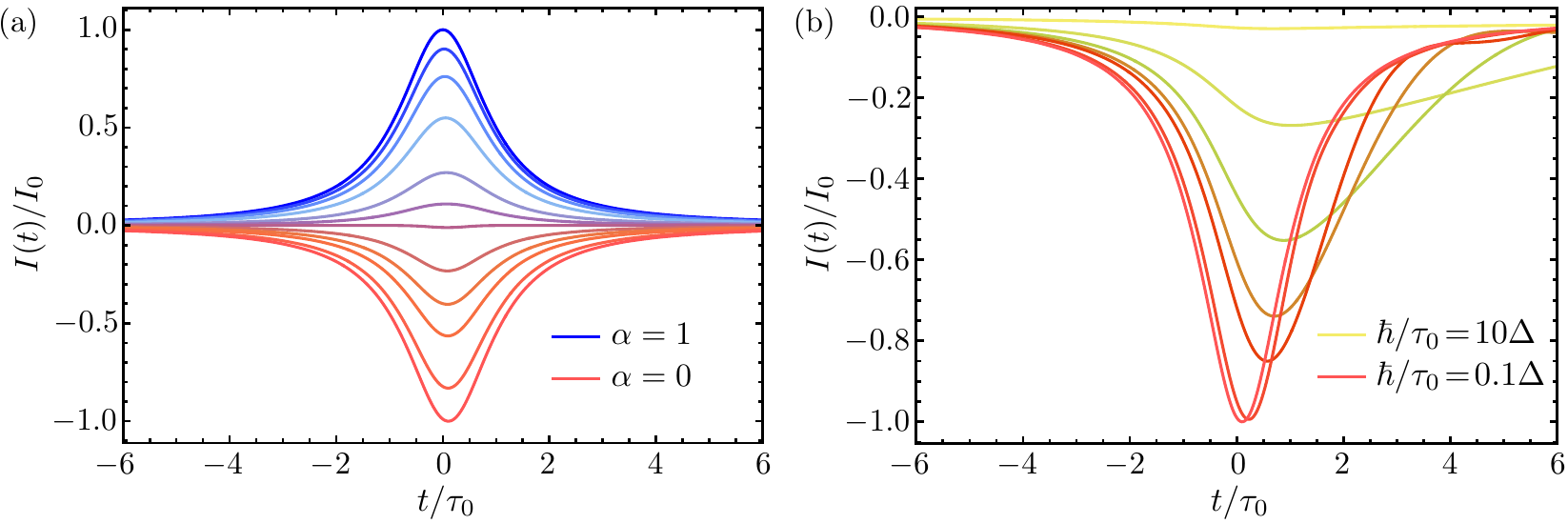}
	\caption{\label{fig:andreev-ref} 
		Andreev conversion of a charge pulse on a superconductor. 
		(a) The time-dependent current, $I(t)$, in the outgoing edge state for different degrees of electron-hole conversion, $\alpha=0$ (full conversion), 0.15, 0.25, 0.3, 0.35, 0.42, 0.45, 0.5, 0.6, 0.7, 0.8, 1 (normal reflection) with the excitation energy well inside the superconducting gap, $\hbar/\tau_0=0.1\Delta$, and $T=0$. 
		(b) The time-dependent current for different excitation energies $\hbar/(\tau_0 \Delta)=0.1$, 0.2, 0.5, 0.66, 1, 2, 10 with perfect electron-hole conversion at the superconductor, $\alpha=0$. The current is divided by the maximum of the injected current denoted by $I_0$. 
	}
\end{figure*}

The voltage pulses are applied to the input (on both spin channels) and far from the superconductor, such that~the~Floquet-Nambu scattering matrix reads
\begin{equation}\label{eq:scat-mat-simp}
\hat{S}_{F}(E_{n},E)= \hat{S}(E_{n}) \hat{J}_{n} ,
\end{equation}
where $\hat{S}(E)$ describes the scattering at the interface with the superconductor and has the Nambu structure in \cref{eq:scat-mat}, while  $\hat{J}_{n} = \operatorname{diag}_{N}(J_{n},J^*_{-n})$ contains the Fourier coefficients of the voltage-induced phase factor,
\begin{equation}\label{eq:fourier}
J_n= \int_{0}^{\mathcal{T}} \frac{\mathrm{d}t}{\mathcal{T}} \me^{\mi n\Omega t} \me^{\mi\varphi(t)},\,\, \varphi(t)=\frac{e}{\hbar} \int_{-\infty}^{t}\mathrm{d}t'  V(t').
\end{equation}
Here, we focus on Lorentzian pulses of width $2\tau_0$ and period $\mathcal{T}$, $eV(t)=-2\hbar \tau_0 \sum_{n=-\infty}^{\infty}[(t-n\mathcal{T})^2+\tau_0^2]^{-1}$, which inject exactly one charge per pulse into the circuit.
The Fourier components then read
$J_{n>0}= -2\sinh(\eta) \me^{-n \eta}$ and $J_0= \me^{-\eta}$, where $\eta= \tau_0\Omega$ determines the overlap of the pulses. 
The distinctive feature of these pulses is that no holes are excited by the drive, since $J_{n<0}=0$~\cite{Levitov_2006,Dubois_2013b}. We take the period to be much longer than the pulse width, $\mathcal{T}\gg\tau_0$, so that the individual charge pulses arrive and scatter off the superconductor one at a time.

{\it The Nambu  state.---} \Cref{eq:scat-mat-simp} describes how the electron and hole components of the incoming charge pulses are mixed at the superconductor interface. In general, the scattering matrix, $\hat{S}$, depends on the energy, which complicates further analysis. However, if all energy scales are well within the superconducting gap $(\Delta>0)$, so that $k_BT,\hbar/\tau_0,\hbar\Omega\ll \Delta$, the scattering amplitudes in $\hat{S}$ can be approximated by constants. A detailed calculation then shows that the excess correlation function can be recast into the remarkably simple form~\cite{Burset_long}
\begin{equation}\label{eq:GF-decoup}
		\hat{G} (x,t;x',t') = 
				\sum_{\xi=e,h} \hat{M}_{\xi}(0,0;x',x) G_{\xi}(t',t)
\end{equation}
in terms of the clean electron $(e)$ and hole $(h)$ Green functions $G_{e,h}(t',t)= \pm \Psi^*_{\mp}(t') \Psi_{\mp}(t) / v_F$, with $\Psi_\pm(t) = \sqrt{\tau_0/\pi} / (t \pm \mi\tau_0)$. Here, the scattering at the superconductor is described by the matrix $\hat{M}_{\xi}$ with elements $[\hat{M}_{\xi}(E',E;x',x)]_{\nu\nu'}= [\hat{\phi}^*_{E'}(x')]_{\nu\nu}  [\hat{S}^*(E')]_{\nu\xi} [\hat{\phi}_{E}(x)]_{\nu'\nu'} [\hat{S}(E)]_{\nu'\xi} $, for $\nu,\nu'=e,h$. 

\Cref{eq:GF-decoup} is a central result of our work, and we can use it to gauge the purity of the state for quantum information processing~\cite{Bisognin_2019}.
Fourier transforming \cref{eq:excess} into the frequency domain and defining 
$\hat{\tilde{G}}_{++}(\omega,\omega')= \hat{\tilde{G}}(\omega>0,\omega'>0)$ for positive frequencies, an important characterization is the purity condition~\cite{Roussel_2021,Burset_long}
\begin{equation}
	v_F\int_0^{\infty} \! \frac{\md \omega}{2 \pi}
		\hat{\tilde{G}}_{++}(\omega_1,\omega) \hat{\tilde{G}}_{++}(\omega,\omega_2)
	= \hat{\tilde{G}}_{++}(\omega_1,\omega_2) .
\end{equation}
If this condition is fulfilled, the state can be understood as a stream of pure electron-hole superpositions, realizing the superconducting equivalent of a perfect single-electron source.
Using \cref{eq:GF-decoup}, we can show that this condition is satisfied for the scattered state as long as it is valid for the incoming state.
In fact, it turns out that the purity condition is fulfilled if all excitation energies are smaller than the superconducting gap, $\hbar\Omega,\hbar/\tau_0 \leq \Delta$, such that no quasi-particles are transferred into the superconductor~\footnote{
Since the sum in \cref{eq:scat-ops} runs up to infinity, there is always a nonzero probability of exciting particles over the superconductor gap $\Delta$. However, the probability amplitudes $J_n$ in \cref{eq:fourier} decay exponentially with the drive frequency $\Omega$ and the pulse half-width $\hbar/\tau_0$. When these quantities are smaller or comparable to $\Delta$ the probability of exciting particles over the gap is negligible~\cite{Burset_long}
}. 
In turn, by scattering single electrons with energies well within the gap off a superconductor, we obtain a stream of pure electron-hole superpositions.
For larger energies, the periodic drive can excite quasi-particles to energies above the gap such that they can be transmitted into the superconductor, reducing the purity of the outgoing state in the edge channels.

{\it Andreev conversion.---} The conditions for producing a pure state are met with superconductors like Al or NbTi, with $\Delta/k_B\simeq \SI{1}{\kelvin}$ and $\Delta/k_B\simeq \SI{10}{\kelvin}$, and with voltage pulses in the gigahertz regime applied to an electronic reservoir at around $\SI{20}{\milli\kelvin}$~\cite{Dubois_2013}. Moreover, irradiated superconducting atomic contacts with the required parameters have been realized~\cite{Urbina_2006}, and their extension to quantum Hall systems is within experimental reach~\cite{Lee_2017,Vignaud2023}. Thus, the scattering of charge pulses on a superconducting contact seems experimentally feasible. Our main prediction is that a clean charge pulse with an energy inside the superconducting gap may undergo a (partial) Andreev conversion and become a coherent superposition of an electron and a hole. The electron and hole content is determined by the scattering amplitudes, but at temperatures below the gap, no other quasiparticles are excited from the Fermi sea. For complete Andreev reflections, the junction would become a perfect electron-hole converter for single-particle pulses, a device application with no equivalent in quantum optics with photons~\footnote{Perfect Andreev conversion always takes place at low energy for triplet superconductors.}. 

To go beyond the expression in \cref{eq:GF-decoup}, we need to describe the energy-dependence of the superconducting scatterer. 
The contact between the edge states and the superconductor is represented by the matrix~\cite{BTK,Beenakker_2014,Burset_long}
\begin{equation}\label{eq:smatrix_SC}
	\hat{S}(E) = \frac{1}{ \me^{\mi\gamma}-\alpha^2 \me^{-\mi\gamma} }\begin{pmatrix}
		2\mi \alpha \sin\gamma & 
		\left( 1 - \alpha^2 \right) \me^{\mi\phi} \\ 
		\left( 1 - \alpha^2 \right) \me^{-\mi\phi} & 
			2\mi \alpha \sin\gamma
\end{pmatrix} ,
\end{equation}
which is an extension of the model in Ref.~\cite{BTK} for chiral channels, where $\phi$ is the superconducting phase and $\cos\gamma=E/\Delta$. 
The electron-hole mixing depends on microscopic details of the contact, like the ratio of the width of the superconductor over the superconducting coherence length~\cite{Beenakker_2011,Beenakker_2014,Clarke_2014,Nazarov_2017,Nazarov_2019,Idrisov_2020,Akhmerov_2022,Klinovaja_2022,Kurilovich2023,Schmidt_2023,Oreg_2023,Balseiro_2023,Baba2025}, which we describe by the parameter $\alpha$ with $\alpha = 0$ ($\alpha = 1$) for perfect Andreev (normal) transmissions. 
Inserting \cref{eq:smatrix_SC} into \cref{eq:current}, we can evaluate the current in the output after the superconductor, which we show in \cref{fig:andreev-ref}(a) for different degrees of electron-hole mixing and excitation energies well inside the superconducting gap. Here, we find that the full microscopic calculation is well captured by \cref{eq:GF-decoup}. 
Indeed, by combining \cref{eq:current,eq:GF-decoup}, the current can be written in terms of a positive and a negative pulse as
\begin{equation}
	I(t) = (2P-1) G_0V(t),
\end{equation}
where $G_0=2e^2/h$ and $P=|S_{eh}(E=0)|^2=(1-\alpha^2)^2/(1+\alpha^2)^2$ is the Andreev conversion probability. 

For $\alpha = 0$, an incoming electron is perfectly converted into a hole, while it is fully reflected off the superconductor for $\alpha = 1$. For values in between, the outgoing state is a coherent superposition of an electron and a hole. In particular, for $\alpha \simeq 0.41$, where $P=1/2$, a charge-neutral pulse is generated by the partial Andreev reflection on the superconductor. The complete Andreev conversion is robust even if the condition $\hbar/\tau_0\ll\Delta$ is relaxed as demonstrated in \cref{fig:andreev-ref}(b), where we increase the excitation energy and observe a clearly negative current even for $\hbar/\tau_0\simeq\Delta$. The figure also shows that the photo-assisted transmissions into the superconductor only have an important contribution for $\hbar/\tau_0\gtrsim\Delta$. In turn, the purity of the outgoing state can be maintained beyond the experimentally most relevant regime of $\hbar/\tau_0\ll\Delta$. This result even holds at finite but low temperatures, where the thermal coupling between the single-particle states and the Fermi sea can safely be neglected~\cite{Moskalets_2018,Ryu2022}. 

\begin{figure}
	\includegraphics[width=1.00\columnwidth]{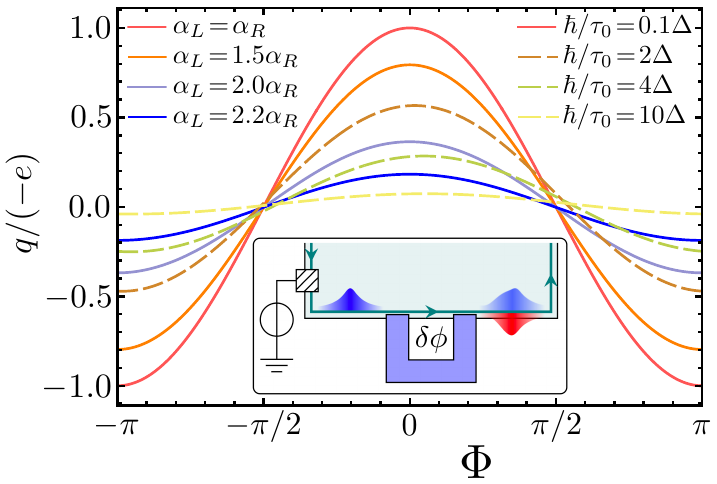}
	\caption{\label{fig:andreev-interf} 
		Tunable electron-hole conversion. The degree of conversion can be controlled by the phase $\Phi$. The average charge per pulse is shown as a function of $\Phi$ for a symmetric setup with $\alpha_L=\alpha_R$ at different drive frequencies (dashed lines), and for a drive $\hbar/(\tau_0\Delta)=0.1$ for different asymmetric configurations (solid lines). Here, we have taken $\alpha_L=0.41$.
	}
\end{figure}

{\it Tunable converter.---} In the setup above, the degree of electron-hole conversion is fixed by the microscopic properties of the superconductor interface. However, we can extend the setup to enable a tunable degree of electron-hole conversion. To this end, we include a second superconductor as in \cref{fig:andreev-interf}~\cite{Zulicke_2005,Ustinov_2007,Schonenberger_2012,Amet_2016,Lee_2017,Ren_2017,Das_2018,Xu_2019}. In this case, the total scattering amplitude depends on the relative phase between the superconductors, $\delta\phi = \phi_{R} - \phi_{L}$, which can be controlled by the external magnetic field
~\cite{[{An interesting alternative would be to use a single superconductor whose scattering amplitudes can be controlled by local gating of the edge states, see }] Trauzettel_2019}. The scattering matrix for the combined system is a product of scattering matrices for each part, $\hat{S}(E,d,\delta\phi) = \hat{S}_R(E,\alpha_{R},\phi_R) \hat{S}_{0}(d) \hat{S}_L(E,\alpha_{L},\phi_L)$, where $\hat{S}_{0}(d) \!=\! \mathrm{diag}_N(\me^{-\mi k_{F}d},\me^{\mi k_{F}d})$ describes the propagation of electrons and holes over the distance $d$ between the superconductors with scattering matrices given by \cref{eq:smatrix_SC}. 
Defining the phase $\Phi=\delta\phi+2k_F d$, we find a simple expression for the electron-hole amplitude at zero energy,
\begin{equation}
	|S_{eh}(E=0)|^2 = P_{LR} + P_{RL} + 2 \sqrt{P_{LR}P_{RL}}\cos \Phi,
	\label{eq:conversionamp}
\end{equation}
with $P_{AB}= P_A(1-P_B)$ and $P_{L/R}$ the Andreev conversion probability at each superconductor~\footnote{Note that, as long as transport is coherent, the distance $d$ between the superconductors only causes a phase shift.}. 
The cosine is an interference term that arises because electrons and holes pick up different phases as they scatter off the superconductors and propagate between them. \Cref{fig:andreev-interf} shows the charge per pulse $q= \int_\text{pulse} \mathrm{d}t I(t)$ as a function of $\Phi$ with the sign and magnitude reflecting the proportion of electrons and holes. For symmetric configurations with $P_L=P_R$, \cref{eq:conversionamp} reduces to $|S_{eh}(0)|^2 = 4P_{LL}\cos^2 \Phi/2$. Then, if the normal and Andreev scattering probabilities at each superconductor are approximately the same ($P_L\sim1/2$), the phase determines if the outgoing superposition is a perfect Andreev converted hole ($\Phi=\pi$), a charge neutral pulse ($\Phi=\pi/2$), or an electron state ($\Phi=0$). 

Our results assume clean samples of typical sizes that are greater but comparable to the superconducting coherence length. Indeed, recent experiments based on encapsulated graphene have separately measured non-superconducting leviton interferometry with coherence lengths of several micrometers~\cite{Assouline:2023,Chakraborti2025} and chiral supercurrents using superconducting contacts with coherence lengths of tens of nanometers~\cite{Vignaud2023}. 
Interactions are known to affect the propagation and shape of the pulses~\cite{Ferraro_2014,Cabart_2018}, but usually at larger scales~\footnote{Interactions may influence the propagation along the edge channel and cause deformations of the pulses. The incoming pulse can be reshaped, so that it is Lorentzian, once it scatters off the first superconductor. By contrast, the electron-hole superposition that it generated next may decohere over a length scale that we estimate to be on the order of $l_\text{dec} \simeq v_{\text{int}} \tau_0$, where the interaction-dependent velocity $1/v_{\text{int}} = 1/v_s - 1/v_c$ is given by the velocity of the spin ($v_s$) and the charge ($v_c$) modes that arise because of interactions~\cite{Ferraro_2014,Cabart_2018}. Typical experiments have $l_\text{dec} \gtrsim \SI{1}{\micro\meter}$~\cite{Freulon_2015,Marguerite_2016,Assouline:2023,Chakraborti2025}, which can be increased using appropriate sample design~\cite{Altimiras_2010,Huynh_2012}. In short, for $d\ll l_\text{dec}$, the electron-hole conversion can be controlled by the magnetic flux. By contrast, for $d\simeq l_\text{dec}$, the reduced visibility of the interferometric signal can be used to measure the interactions along the edge~\cite{Roussel_2017,  Martin_2023b}}. 
We describe some of these effects in our companion paper~\cite{Burset_long}.

{\it Conclusions and outlook.---} We have proposed and analyzed a setup for the on-demand electron-hole conversion of charge pulses by Andreev reflections on a superconductor and thereby designed an important building block for future electron quantum optics experiments with no photonic counterpart. To this end, we have developed a Floquet-Nambu formalism that can describe the Andreev reflections of charge pulses with the full technical details provided in a companion paper that also includes an in-depth discussion of the superconducting pairing potential~\cite{Burset_long}. For realistic experimental conditions with low excitation energies and temperatures compared to the superconducting gap, an incoming charge pulse can be converted into a coherent superposition of an electron and a hole with the degree of conversion controlled by the magnetic flux in an interferometric setup. 
This setup constitutes an on-demand source of flying qubits where the tunable electron-hole superposition can function as a carrier of quantum information~\cite{sequoia2022}. As future work, one may investigate interferometry using such a source and a known signal for quantum state tomography or even collisions between two flying qubits~\cite{Chakraborti2025}. 

\acknowledgments
{\it Acknowledgments.---} We thank A.~A.~Clerk and J.~Keeling for valuable discussions. 
The work was supported by Spanish CM ``Talento Program'' project No.~2019-T1/IND-14088 and No.~2023-5A/IND-28927, the Agencia Estatal de Investigaci\'on project No.~PID2020-117992GA-I00, No.~PID2024-157821NB-I00 and No.~CNS2022-135950 and through the ``María de Maeztu'' Programme for Units of Excellence in R\&D (CEX2023-001316-M), the CSIC/IUCRAN2022 under Grant No.~UCRAN20029, and Research Council of Finland through the Finnish Centre of Excellence in Quantum Technology (project~number~352925).


%

\end{document}